\documentclass[english,twocolumn]{revtex4-2}
\usepackage[T1]{fontenc}
\usepackage[latin9]{inputenc}
\usepackage{float}
\usepackage{mathrsfs}
\usepackage{graphicx}
\usepackage{babel}
\begin{document}
\title{Input-driven circuit reconfiguration in critical recurrent neural
networks. }
\author{Marcelo O. Magnasco}
\address{Laboratory of Integrative Neuroscience, Rockefeller University, New
York}
\email{mgnsclb@rockefeller.edu}

\begin{abstract}
Changing a circuit dynamically, \emph{without actually changing the
hardware itself,} is called reconfiguration, and is of great importance
due to its manifold technological applications. Circuit reconfiguration
appears to be a feature of the cerebral cortex, and hence understanding
the neuroarchitectural and dynamical features underlying self-reconfiguration
may prove key to elucidate brain function. We present a very simple
single-layer recurrent network, whose signal pathways can be reconfigured
``on the fly'' using only its inputs, \emph{with no changes to its
synaptic weights}. We use the low spatio-temporal frequencies of the
input to landscape the \emph{ongoing activity}, which in turn permits
or denies the propagation of traveling waves. This mechanism uses
the \emph{inherent properties} of dynamically-critical systems, which
we guarantee through unitary convolution kernels. We show this network
solves the classical \emph{connectedness }problem, by allowing signal
propagation only along the regions to be evaluated for connectedness
and forbidding it elsewhere. 
\end{abstract}
\maketitle
The brain changes its own hardware at different timescales, for example
through synaptic plasticity, synaptic remodeling and adult neurogenesis
\citep{plasticity}, and thus retains an extraordinary amount of long-term
plasticity. But also, some brain areas evidence the ability to switch
between several different functions depending on context, and these
switches happen at faster speeds that seem compatible with a physical
change in the underlying neural substrate \citep{switching,switching2}.
Hence it is now appreciated that \emph{the brain dynamically reconfigures
its operations without physical reconfiguration} \citep{cosb17}.
One potential mechanism that has been strongly implicated in such
rapid reconfigurations is \emph{ongoing brain activity} \citep{ongoing},
for example ongoing traveling waves \citep{travelingwaves}. Hence,
as advocated in \citep{cosb17} it is fundamental to understand the
dynamical principles through which a complex network can reconfigure
its operations, by managing and curating its own dynamical state. 

I shall show below that critical dynamics, through its \emph{inherent}
\emph{singular dependence} of the dynamical state on the input, allows
a network to be reconfigured rapidly and efficiently by suitable choice
of inputs. The inputs to such a network conceptually contain both
external sensory information as well as instructions from other brain
areas. Information flows through such a network along pathways which
can be dynamically changed, with no changes to the synaptic weights,
and without any ``special circuitry'' being involved. I shall construct
an example, the simplest I could manage, where the input contains
both control and signaling components in separate spatiotemporal bands.
The control component (low spatial and temporal frequencies) will
\emph{literally} be an image of walls and channels; the signaling
component (higher temporal frequencies) will consist of point-wise
signal sources generating traveling waves; these waves will freely
propagate through the channels but will exponentially attenuate into
the walls, or even reflect off of them. As they do, they will propagate
into every place that's allowed by the input, implementing de facto
a floodfill (``paint bucket'') algorithm. This relates to a classical
problem: how do our brains \emph{detect connectedness} of areas in
visual stimuli \citep{floodfill1,floodfill2}, a problem which is
arguably not solvable by feedforward architectures of fixed depth.
It is argued cogently in \citep{floodfill3} that this is in fact
a touchstone for the central role of recurrence in the nervous system. 

I will use a single-layer recurrent convolutional neural network with
unitary (critical) coupling kernel. The family of such networks was
derived as a Poincaré map from critical ODEs in \citep{chaos}, and
their basic dynamical analysis is outlined in \citep{arxiv} and \citep{chandramm}.
The state is given by a complex-valued ``neural'' layer $Z$, henceforth
a lattice of complex numbers in either 1 or 2 dimensions with periodic
boundary conditions. Its evolution in time is denoted as \emph{$Z_{n}$}
where $n$ is (integer) time, through the recursion
\begin{equation}
Z_{n+1}=\phi\left(U\otimes Z_{n}+I_{n}\right)\label{eq:curnn}
\end{equation}
where $\otimes$ is the convolution operation natural to $Z$; $U$
is a (unitary) convolution kernel; $I_{n}$ is a sequence of external
inputs of the same shape as $Z$; and $\phi$ is a smooth, scalar
``activation function'' operating element-wise on $Z.$ A natural
choice of $\phi$ for this construction is $\phi(z)=z/\sqrt{1+|z|^{2}}$,
a complex-valued phase-preserving sigmoid \citep{chandramm}. We choose
$U$ to be a \emph{unitary} kernel, i.e. its associated linear operator's
eigenvalues $\lambda$ lie on the unit circle: $|\lambda|=1$. We
generate unitary convolution kernels by convolutional exponentiation
of an anti-Hermitic kernel \citep{arxiv}: 
\begin{equation}
U=e_{\otimes}^{A}\equiv\mathscr{F}^{-1}\left\{ \exp\left(\mathscr{F}\left\{ A\right\} \right)\right\} \label{eq:convexpo}
\end{equation}
where the exponential is element-wise, $\mathscr{F}$ and $\mathscr{F}^{-1}$
the direct and inverse Fourier transforms in the dimension $D$ appropriate
to $Z$ and $K$, and $A^{\dagger}=-A$, i.e. antiHermitic. 

Finally, we remind the reader that the eigenvalues of a convolution
with a kernel $U$ are the individual elements of the Fourier transform
of $U$, and the corresponding eigenvectors are the Fourier modes
associated to those elements. The Fourier transform of the kernel
$A$ giving rise to $U$ is is called the \emph{dispersion relation
}of the kernel; since it is purely imaginary, it associates a non-decaying
frequency to each wavenumber. 

\begin{figure}[tb]
\includegraphics[width=8cm]{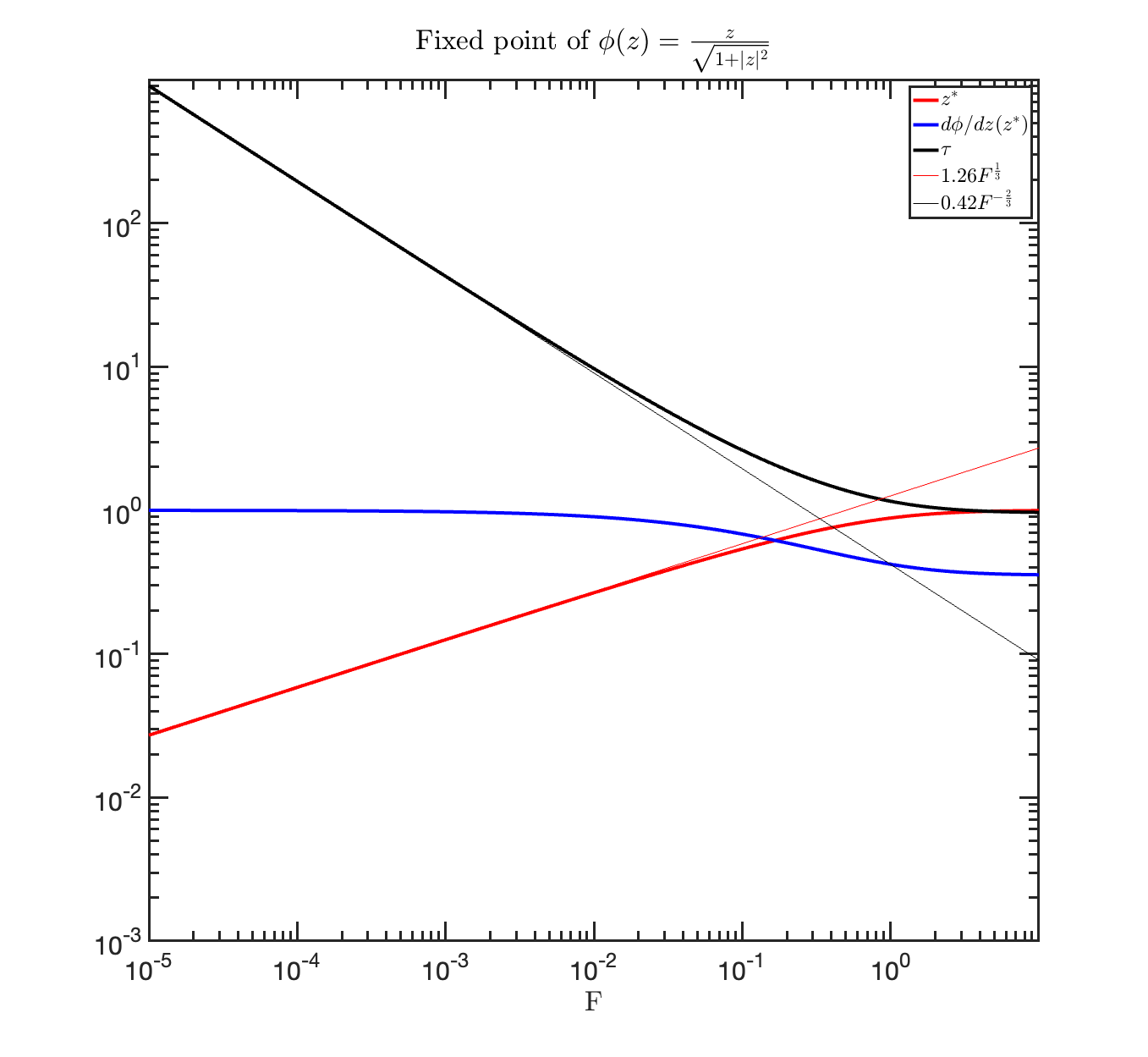}\caption{From \citep{chandramm}. The fixed point $z_{*}=\phi(z_{*}+I)$ with
$\phi(z)=z/\sqrt{1+z^{2}}$; $z_{*}$ as a function of $I$ (red),
together with $\phi'(z_{*})=\left(1+z_{*}^{2}\right){}^{-\frac{2}{3}}$,
the slope of $\phi$ at the fixed point controlling decay rate (blue),
and $-1/\log(\phi'(z_{*}))$ (black), the relaxation time to the fixed
point. The thin lines are power-law asymptotics to the low $I$ regime:
$z_{*}\approx\left(2I\right)^{\frac{{1}}{3}}$ and $\tau\approx\frac{2}{3}\left(2I\right)^{-\frac{{2}}{3}}$.
\label{fig:onedfixedpoint}}
\end{figure}

The propagation of perturbations through this system can be analyzed
in terms of linear stability of dynamical systems. The derivative
of the state $Z_{n}$ at time $n$ wrt the initial state $Z_{0}$
at time $0$ telescopes through the chain rule into a product of derivatives
\citep{devaney,strogatz}, each one of which contains two terms: the
constant unitary matrix representing the action of the convolution
with $U$ (denoted by $[U\otimes]$ in \citep{arxiv}), times the
matrix of derivatives of $\phi$ evaluated at its argument at the
appropriate time-step, which, since $\phi$ is element-wise, is strictly
diagonal: 
\[
\frac{\partial Z_{n}}{\partial Z_{0}}\approx\prod_{i=0}^{n-1}\Gamma_{i}\cdot[U\otimes]\quad{\rm where}\quad\Gamma_{i}\doteq{\rm diag}\left[\phi'(MZ_{i}+I_{i})\right]
\]
If $\phi$ is a sigmoidal function, and satisfies $\phi'(0)=1$ and
$0\le\phi'(x\ne0)<1$, then for $I=0$: (a) the state $Z\equiv0$
is globally attracting (i.e. all activities in the layer decay to
$0$ regardless of initial conditions), and (b) the system is ``critical''
in that the derivative above is a power of the constant unitary matrix
and thus unitary itself. Unitarity of the coupling matrix has manifold
other implications beyond the dynamical ones we explore here. In particular
unitary couplings have been employed in the ANN literature \citep{unitaryrnns}
as a means to overcome the vanishing gradient problem \citep{vanishing}.
The fact that the same matrix $\Lambda\cdot[U\otimes]$ controls both
asymptotic stability as well as back-propagation is a connection that
will be explored elsewhere. 

For $\phi(z)=z/\sqrt{1+|z|^{2}}$ the response to input forcing at
the same frequency as an eigenvalue of $K$ results, as shown in \citep{chandramm},
in the singular nonlinear cubic-root compression characteristic of
the forced Hopf bifurcation \citep{choi1998,eguiluz2003} wherein
the amplitude of the response is proportional to the cubic root of
the amplitude of the input when driven exactly at resonance.b We will
illustrate in a scalar example:
\[
z_{n+1}=\phi(z_{n}+I_{n})=\frac{(z_{n}+I_{n})}{\sqrt{1+\left(z_{n}^{2}+I_{n}\right)}}
\]
where $z$ is now a single complex number. For constant $I$, this
recurrence converges to a fixed point given by $\left(z_{*}+I\right)^{2}=\frac{z_{*}^{2}}{1-z_{*}^{2}}$
from where asymptotically $z_{*}=(2I)^{1/3}$. Perturbations around
$z_{*}$ evolve, to linear order, through powers of $\phi'(z_{*}+I)$:
\[
z^{*}+\delta_{n+1}=\phi(z^{*}+I+\delta_{n})\approx\phi(z^{*}+I)+\phi'(z^{*}+I)\delta_{n}
\]
See Figure \ref{fig:onedfixedpoint}. 

\begin{figure}[t]
\includegraphics[width=8.5cm]{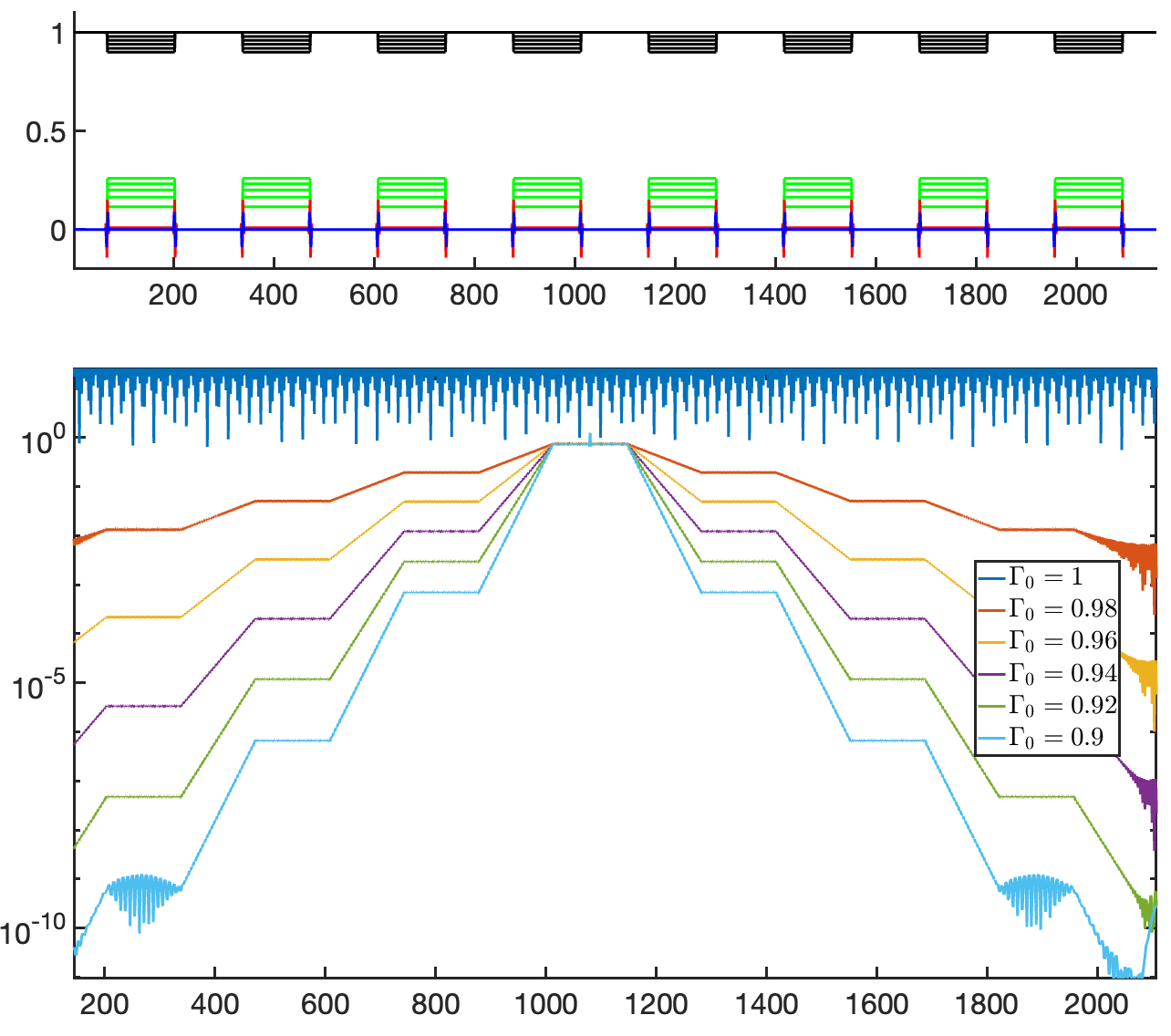}\caption{1D network (N=2048). Top panel. $\Gamma$ (top line, black) was chosen
to have 8 alternating cycles between $\Gamma=1$ and $\Gamma=\Gamma_{0}$.
From there $Z_{*}$ is computed (green line) and then $I_{0}$ (real
part, red; imaginary part, blue). Bottom panel: An oscillatory signal
is injected at the center (a non-attenuating region) and propagates
outwards. Notice it exponentially decays as it traverses $\Gamma<1$
areas and does not attenuate through $\Gamma=1$ areas. \label{fig:spatialattenuation}}

\end{figure}

We now carry this out in a spatially extended system. We first review
the case already considered in \citep{chandramm}: a spatially and
temporally constant input generates a steady state; an oscillatory
perturbation injected into a single site generates waves that propagate
outwards from the injection site and attenuate as they do. The constant
component creates ``ongoing activity'', and the operating point
of the dynamical system shifts to values of $Z$ for which the derivative
of $\phi$ is no longer 1. The oscillatory input then propagates against
this background. 

We write the input as $I_{n}=I_{0}+\alpha\lambda^{n}\delta_{0}$,
where $I_{0}$ is a spatially and temporally constant input, $\lambda=e^{i\theta}$,
and $\delta_{0}$ the image with a single $1$ at element $(0,0)$.
Following standard dynamical-systems perturbation analysis, we will
analyze the behavior around $\alpha\to0$, and so henceforth we assume
$\alpha\ll Z^{*}$. For $\alpha=0$ our system reaches a steady state
given by 
\begin{equation}
Z^{*}=\phi(U\otimes Z^{*}+I_{0})\label{eq:fixpoint}
\end{equation}
where it follows that $Z^{*}$ is spatially and temporally constant.
See also Figure \ref{fig:onedfixedpoint}. 

For $\alpha\ne0$ we write $Z_{n}=Z^{*}+\Delta Z_{n}$ into \ref{eq:curnn}
to obtain 
\[
Z^{*}+\Delta Z_{n+1}=\phi(U\otimes Z^{*}+U\otimes\Delta Z_{n}+I_{0}+\alpha\lambda^{n}\delta_{0})
\]
 and expanding the RHS to first order we obtain 
\begin{eqnarray}
Z^{*}+\Delta Z_{n+1} & = & \phi(U\otimes Z^{*}+I_{0})+\cdots\label{eq:derphiatfixedpoint}\\
 & + & \phi'(U\otimes Z^{*}+I_{0})\left(U\otimes\Delta Z_{n}+\alpha\lambda^{n}\delta_{0}\right)\nonumber 
\end{eqnarray}
the first terms in both the right-hand side and left-hand side cancel,
and calling $\Gamma$ the value 
\begin{equation}
\Gamma=\phi'(U\otimes Z^{*}+I_{0})=\phi'(\phi^{-1}(Z^{*}))\label{eq:Gamma}
\end{equation}
we reach a linear propagation equation for the perturbation generated
by the oscillatory input: 
\begin{equation}
\Delta Z_{n+1}=\Gamma\cdot\left\{ U\otimes\Delta Z_{n}+\alpha\lambda^{n}\delta_{0}\right\} \label{eq:perturbations}
\end{equation}
from where we see that the wave generated by the oscillatory input
attenuates by a factor of $\Gamma$ every time-step. This temporal
attenuation then transforms into a spatial attenuation due to the
finite speed of propagation of the wave (its group velocity): the
slower the wave, the many more factors of $\Gamma$ are incurred in
traversing a given distance. As shown in \citep{chandramm} the asymptotic
state for this equation can be computed in closed form. Writing $\Delta Z_{n}=R^{*}\lambda^{n}$
and substituting in the equation above we get $R^{*}\lambda=\Gamma(U\otimes R^{*}+\alpha\delta_{0})$,
an equation which is explicitly solvable in Fourier space
\[
R^{*}=\mathscr{F}^{-1}\left[\cdot\frac{\alpha\mathbf{\mathscr{F}[\delta_{0}]}}{\lambda/\Gamma-\mathscr{F}[U]}\right]
\]
where $\cdot-$ denotes pointwise division. For $\Gamma\to1$ as $\lambda$
approaches an eigenvalue of $U$ (these are, in fact, the elements
of $\mathscr{F}[U]$ and are, as discussed arranged on the unit complex
circle) the denominator has a pole and the corresponding eigenvector
dominates $R^{*}$, but for $\Gamma<1$ the denominator is bounded
away from $0$ and the responses are spatially-decaying exponentials
around the injection point. 

\begin{figure}[t]
\includegraphics[width=4.2cm]{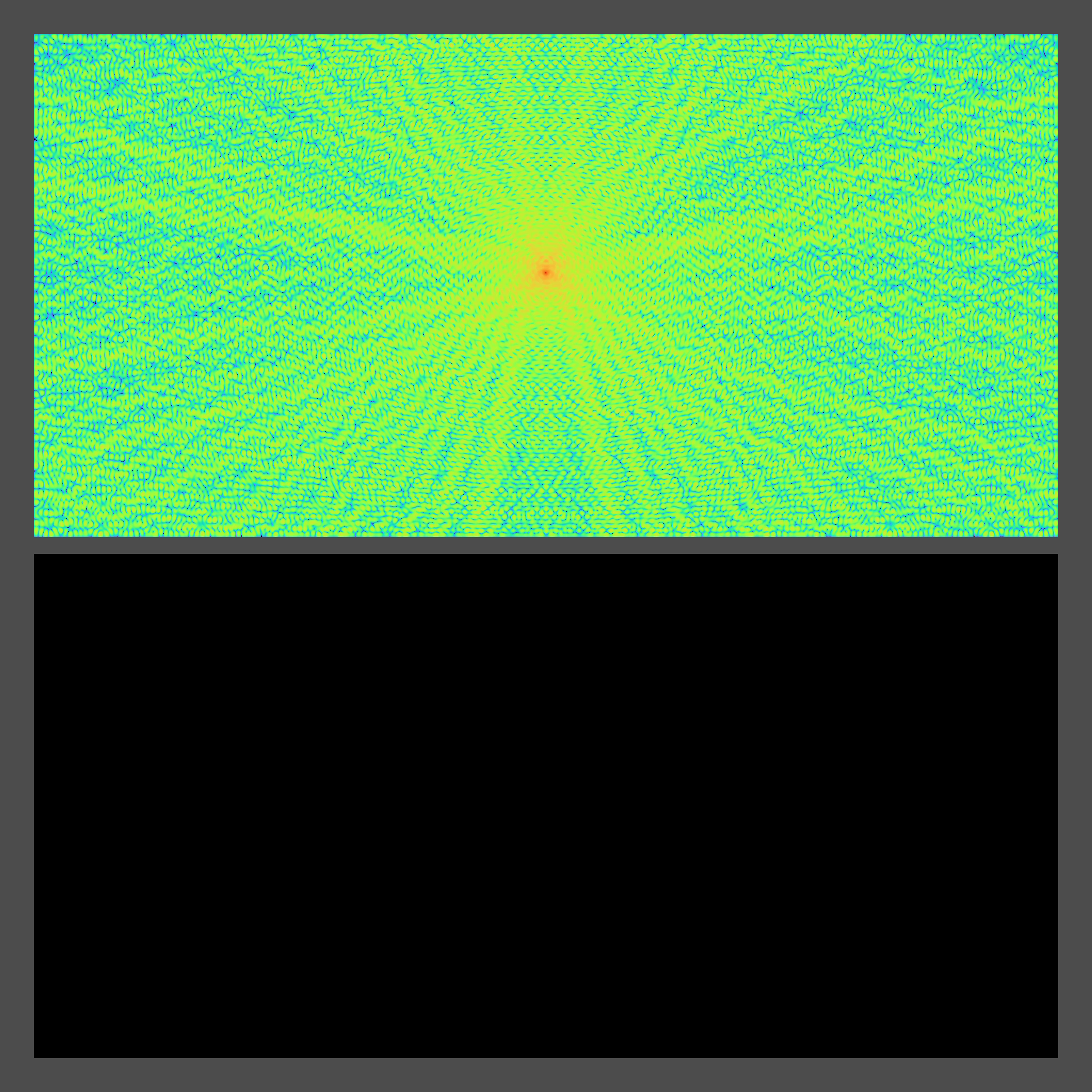} \includegraphics[width=4.2cm]{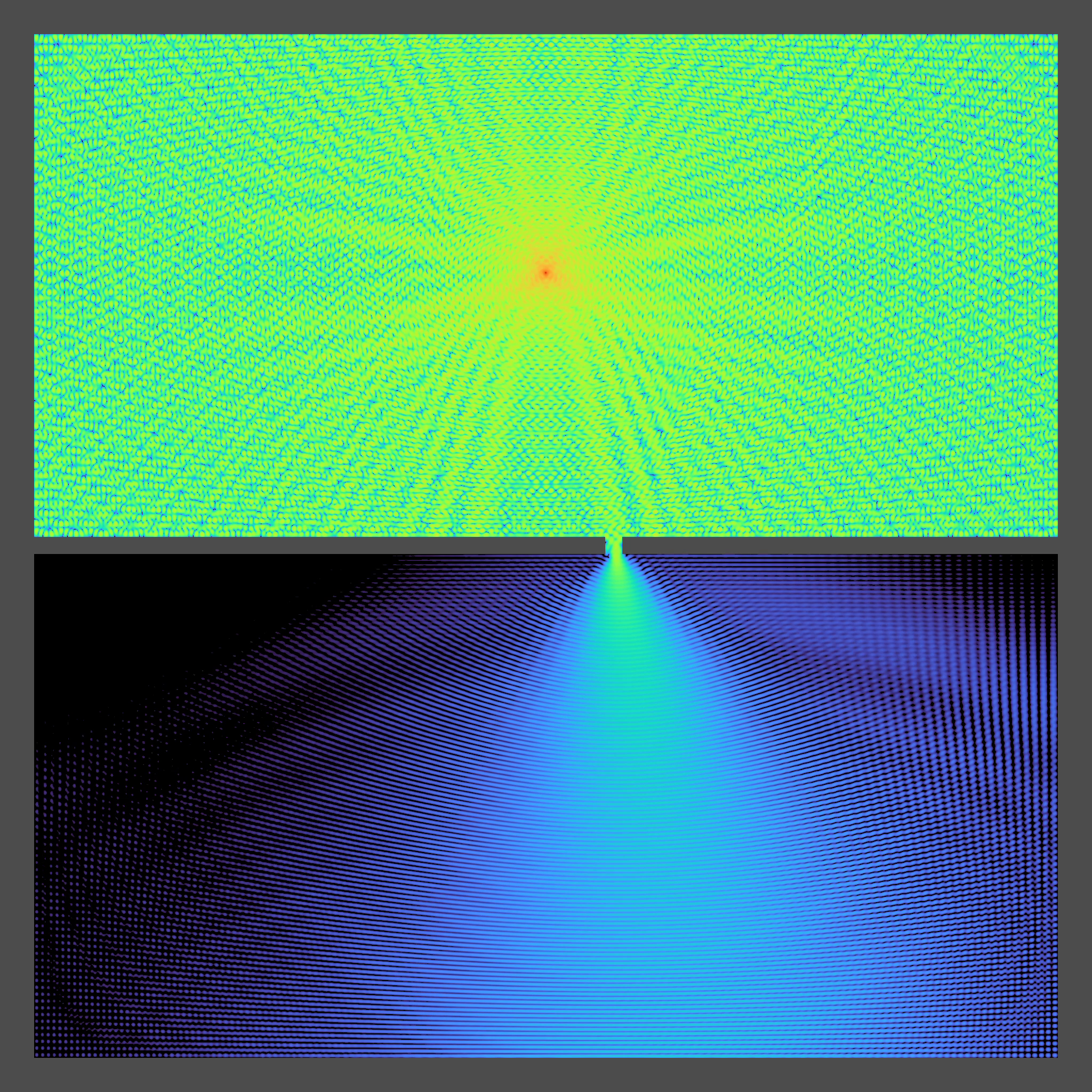}

\caption{Two vertically-stacked boxes. The gray areas cannot be traversed.
In addition to the input $I_{0}$ patterning the boxes, an additional
oscillatory input is applied at the center of the top box. Left, the
middle wall is intact; the signal injected at the top box stays in
the top box. Center, a small aperture is broken in the middle wall;
the signal leaks into the bottom box. Full movies in Supp Mat. This
demonstrates the walls act as a geometric IF statement: something
is allowed to happen or not depending on an input. \label{fig:twoslits}}

\end{figure}

\begin{figure}[H]
\includegraphics[width=8.2cm]{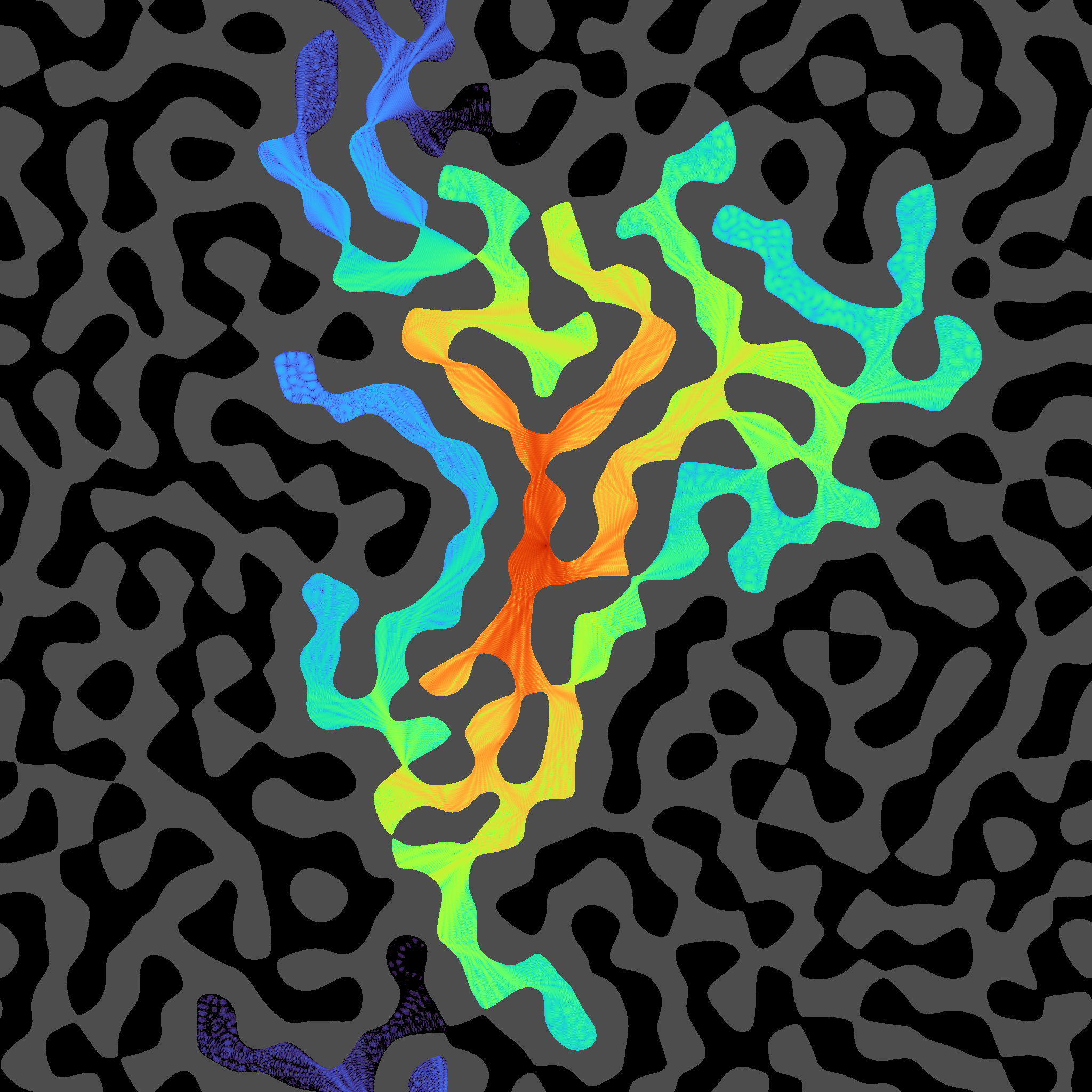}\caption{A labyrinthine pattern was created using band-passed noise to pattern
areas where propagation is allowed ($\Gamma=1$, in black) or strongly
disallowed ($\Gamma=0.01$, gray). An oscillating signal at the frequency
of an eigenvalue was then injected at a single pixel the center of
the figure; the network was evolved and this figure shows $\log(|Z|)$
in color code. $Z$ is a 2048x2048 long complex array; $U$ is the
exponential of a numerical Laplacian kernel (times $i$). Please notice
that the oscillatory signal did not fill non-contiguous areas, implementing
the classic \emph{floodfill }algorithm (a.k.a. \emph{paint bucket}).
Full movies are in Supplementary Materials \label{fig:Floodfill}}
\end{figure}

\begin{figure}[h]
\includegraphics[width=4cm]{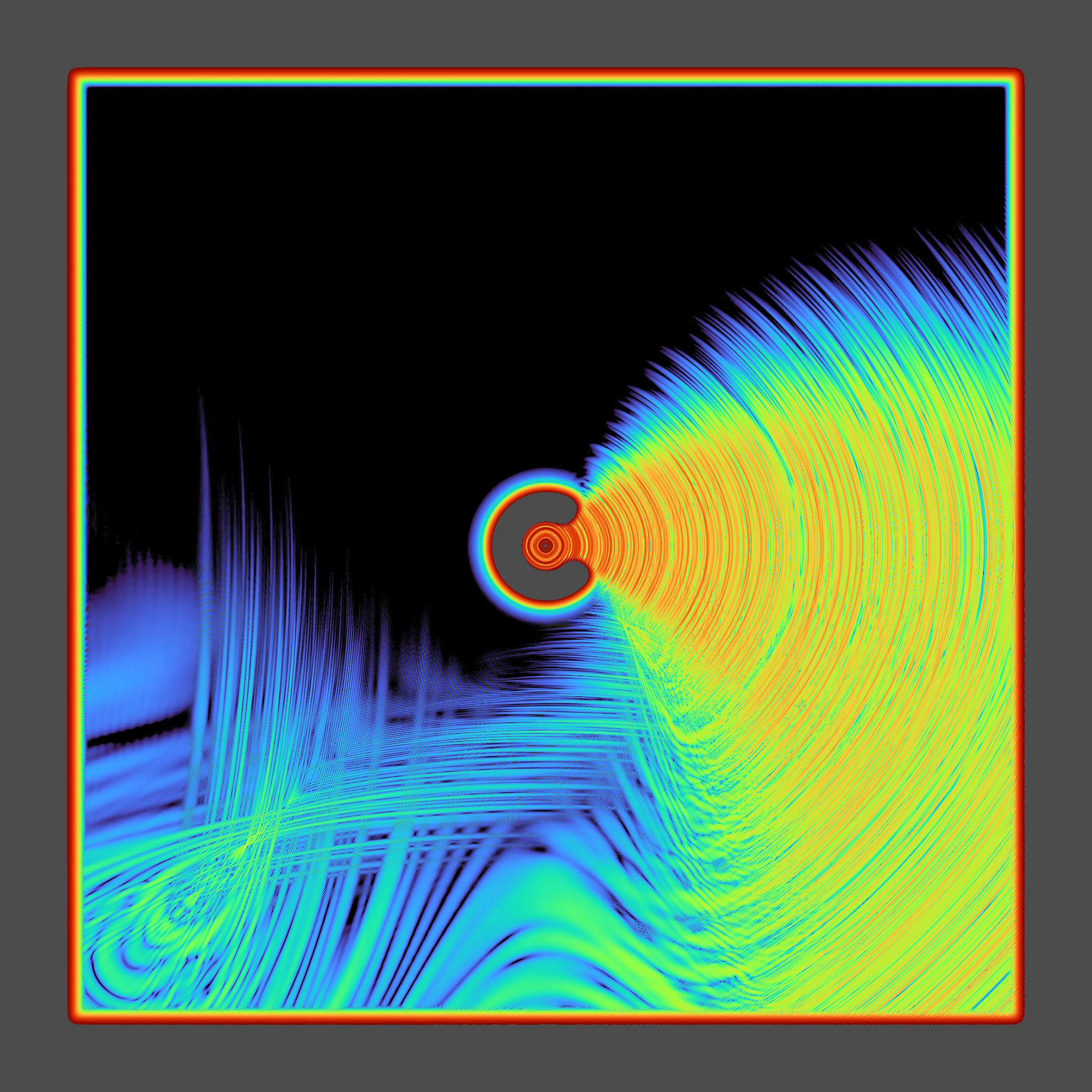}\includegraphics[width=4cm]{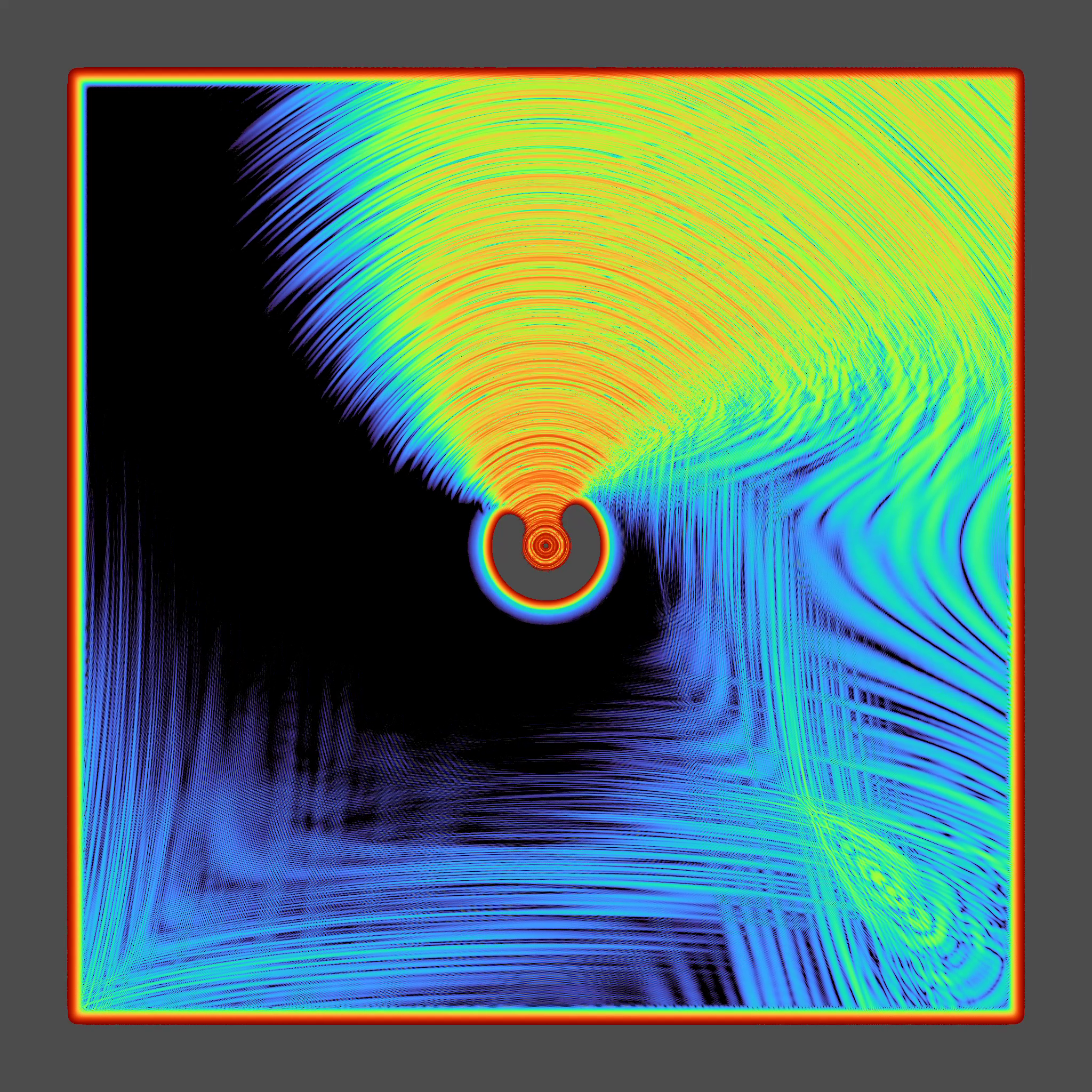}

\includegraphics[width=4cm]{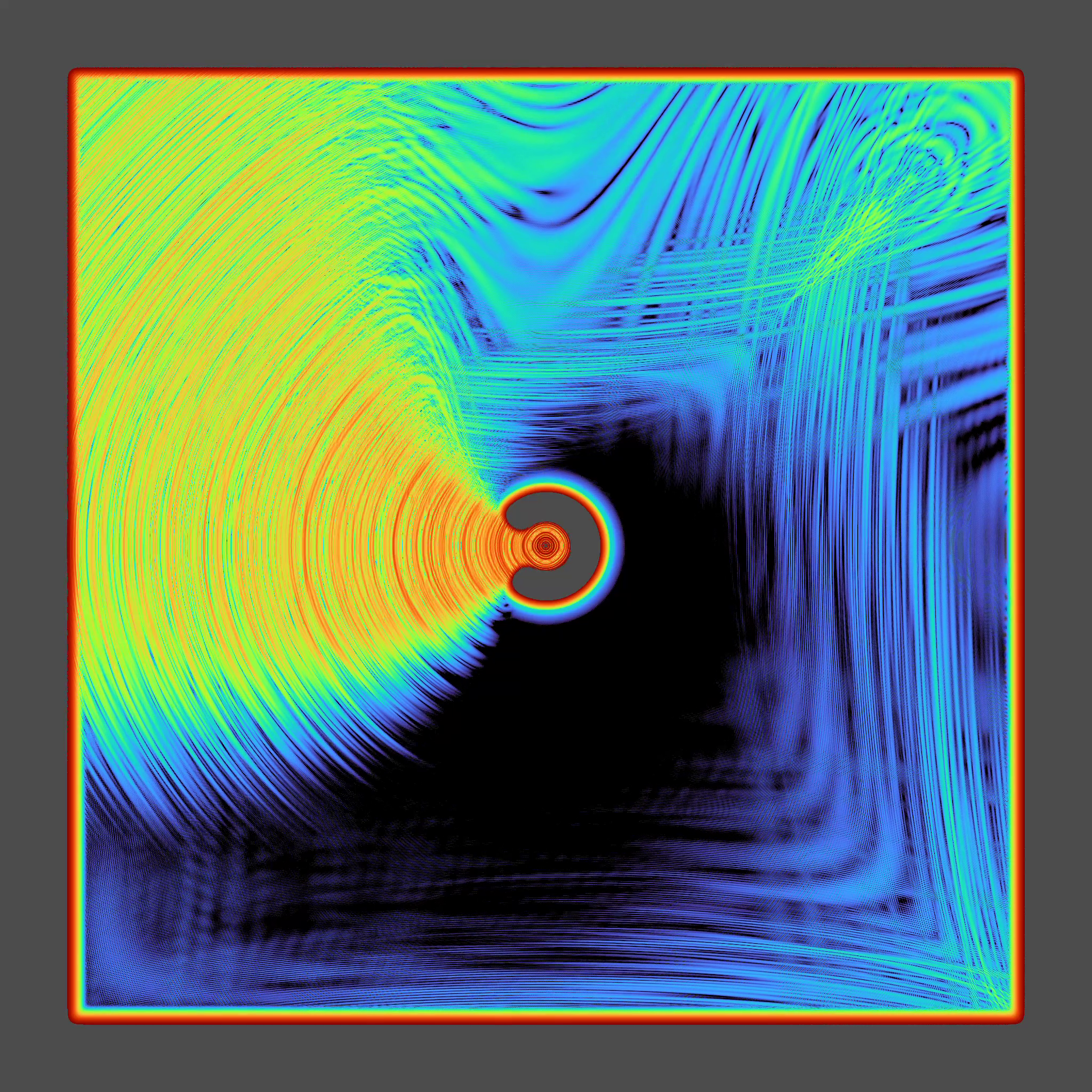}\includegraphics[width=4cm]{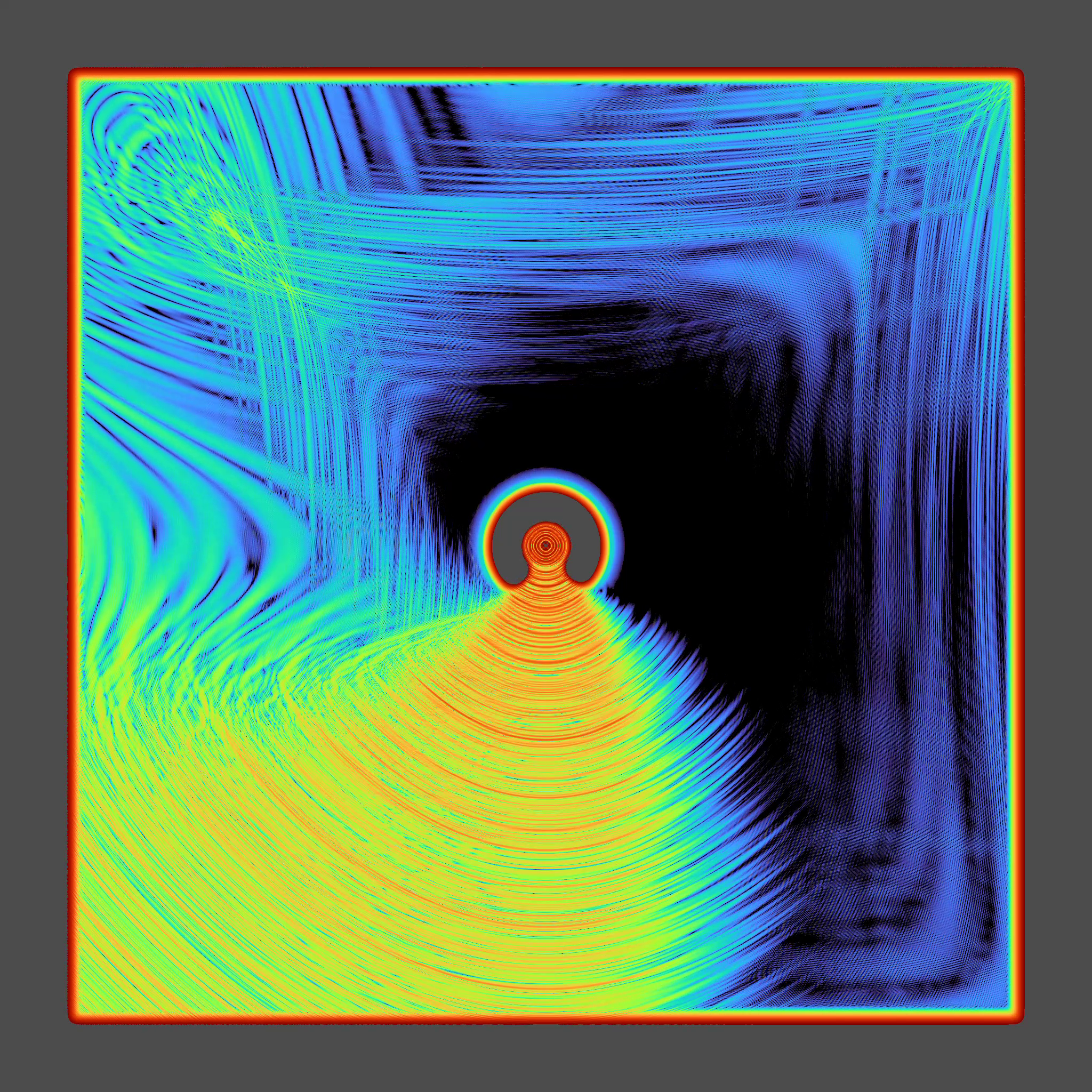}\caption{Lighthouse. \label{fig:Lighthouse}}
\end{figure}

The previous discussion suggests to use a spatially varying $I_{0}$
to \emph{spatially} pattern areas where a signal is not allowed to
enter and others where it is free to propagate. Given a temporally
constant but spatially-patterned input $I^{*}$, iteration of Eq \ref{eq:curnn}
will reach a spatial pattern of ongoing activity $Z^{*}$ given by
the fixed point Eq \ref{eq:fixpoint}. Perturbations around the fixed
point evolve according to Eq \ref{eq:perturbations}, through the
derivative of the activation function $\phi$ evaluated at its arguments
$U\otimes Z^{*}+I_{0}$; this derivative $\Gamma\doteq\phi'(U\otimes Z^{*}+I_{0})$
is now spatially patterned and will generate areas of unobstructed
wave propagation wherever $|\Gamma|=1$, wave absorption whenever
$|\Gamma|<1$, wave amplification and regeneration when $|\Gamma|>1$,
and complex scattering effects when the phase of $\Gamma$ varies
spatially. 

To engineer a given outcome we have to work our way backwards through
the equations: from the desired $\Gamma,$ compute the $Z^{*}$ that
gives rise to $\Gamma$, then compute the $I^{*}$ that gives rise
to $Z^{*}$. Start with the construction of a desired spatially-varying
$\Gamma$: this is an array, the same size as the layer $Z$, and
wherever $|\Gamma|=1$ waves propagate unhindered, and wherever $|\Gamma|<1$
waves attenuate. Then we invert Eq \ref{eq:Gamma} to get the arguments
of $\phi'$ as a function of $\Gamma$. For our choice of $\phi,$
we have $\phi'(z)=\left(1+z^{2}\right)^{-3/2}$ so $z=\sqrt{\phi'^{-2/3}-1}$
and thus 
\[
U\otimes Z^{*}+I_{0}=\sqrt{\Gamma^{-2/3}-1}
\]
This gives us both the ongoing activity and the input as a function
of $\Gamma$. To simplify we note that the entire LHS of this equation
is the argument to $\phi$ in Eq \ref{eq:fixpoint}: apply $\phi$
to both sides and substitute \ref{eq:fixpoint} to solve for $Z^{*}$
only as a function of $\Gamma$: 
\[
Z^{*}=\phi\left(\sqrt{\Gamma^{-2/3}-1}\right)
\]
and then finally use again Eq \ref{eq:fixpoint} to solve for what
the input is that gives rise to a given ongoing activity: $I_{0}=\phi^{-1}\left(Z^{*}\right)-U\otimes Z^{*}$
\begin{equation}
I_{0}=\sqrt{\Gamma^{-2/3}-1}-U\otimes\left[\phi\left(\sqrt{\Gamma^{-2/3}-1}\right)\right]\label{eq:inputfromgamma}
\end{equation}

Equation \ref{eq:inputfromgamma} is our central result. Using this
equation, we can obtain any desired pattern of spatial attenuation. 

To illustrate how input can be used to generate or open barrierts
to passage, in Figure \ref{fig:twoslits}, we used a (complex) layer
$Z$ of size 2048x2048, and $U=e_{\otimes}^{i\Delta}$ with $\Delta$
a numerical Laplacian kernel. An array $\Gamma$ was created defining
two boxes stacked over each other; the walls of these boxes are strongly
attenuating and cannot be traversed by signals, and the $I_{0}$ was
computed. An oscillatory signal was then added to $I_{0}$ applied
at a single pixel at the center of the top box. When the wall between
the boxes is intact, the signal at the top cannot reach the bottom
box. If, on the other hand, a small hole is made in the middle wall,
then the signal can reach the bottom. Therefore, this shows how the
input signal $\Gamma$ can change whether or not other signals are
able to propagate; in effect, this is a geometric IF statement that
allows something to happen or not happen depending on what the input
is. 

In Figure \ref{fig:Floodfill}, we used a complex-valued 2048x2048
layer $Z$; created a labyrinthine pattern for $\Gamma$ (using thresholded
band-pass noise) and computed the $I_{0}$ and $Z^{*}$ that generate
it; then an oscillatory signal was applied at the center of $Z$ and
allowed to propagate. The signal propagates only along the allowed
channels, filling contiguous regions but not invading non-contiguous
regions, thus implementing \emph{floodfill, }a computation that, as
discussed in the introduction, has been shown not to be computable
using feedforward networks \citep{floodfill1,floodfill2}.

To illustrate how to use a \emph{moving boundary condition}, in Figure
\ref{fig:Lighthouse} we used as an input, a slowly rotating image
of $3/4$ of an annulus (i.e. an annulus with one quarter missing)
as our impenetrable wall, and placed a signal source at its center.
As a result, the signal escapes the annulus through the opening, which
then rotates. forming a little ``lighthouse''. This shows input-dependent
directional selection of a signal, or dynamic routing. 

To conclude, we have presented a fairly minimal model of dynamic reconfiguration.
The model is inherently geometrical, because it is laid out on an
Euclidean lattice supporting convolutions, but illustrates concepts
that can be used on more complex underlying topologies: (a) to use
connections that support travelling waves, by use of unitary dynamics,
(b) to exploit the inherent sensitivity of such a system to sculpt
ongoing activity, (c) use that activity to control the space through
which the travelling waves can move in a dynamical fashion. This mechanism
can be viewed as a geometric form of an IF statement: depending on
some data, a computation is carried out, or not. The ways that such
a geometrical construct can be exploited are undoubtedly numerous. 

We make no representation that this model is optimal in any way other
than being conceptually simple, nor we make a claim that it is directly
applicable to computational neuroscience other than illustrating a
dynamical mechanism. No effort was made to optimize any of the available
degrees of freedom: the kernel, the activation function, and the patterns
used as input. It is clear that a designed or optimized network could
do materially better at the tasks presented. The virtue of this model
is one: Figs 3,4, and 5 were run on the \emph{same, identical network}:
same kernel, same activation\emph{. }The only thing that changes between
them is the input $I^{*}$ given to pattern the tasks. (Supplementary
Movie 7 shows dynamics with a random 7x7 kernel, to illustrate the
fact that the results are \emph{not }sensitive to the choice of kernel). 

I would like to acknowledge the late Hector Vucetich, who taught me
how to exponentiate a derivative, as well as the late Mitchell Feigenbaum
with whom I first debated dynamics in neuroscience. I would like to
thank Leandro Alonso for his insights in \citep{chaos} at the genesis
of this line of research; Oreste Piro and Diego Gonzales from whom
I learned the value of Poincare sectioning an ODE, Shaul Druckmann
who highlighted the connections to connectedness, and A. Karuvally,
T. Sejnowsky and H. Siegelman for helpful discussions. 

\pagebreak{}

\section*{Supplementary Material}

\subsection{Supplementary movies}

It's hard to explain without movies, hence virtually any figure here
has movies to match. Here's the list; all movies are labeled SM\%d.
All cuRNN movies are 2048x2048, and in 2-8, in order to show asymptotic
behaviour better, time increases with frame number with a small quadratic
component. The summary movie concatenates all 9 at half resolution
with compression. 
\begin{enumerate}
\item From Fig \ref{fig:onedfixedpoint}. Depiction of the motion of the
fixed point and its slope as a function of the input. 
\item From Fig \ref{fig:twoslits} left panel: zero slit, $U=e_{\otimes}^{i\Delta}$,
\item From Fig \ref{fig:twoslits} center panel: one slit
\item (not shown): two slits 
\item From Fig \ref{fig:Floodfill} left panel: labyrinth floodfill, white
noise, $U=e_{\otimes}^{i\Delta}$
\item (not shown): labyrinth floodfill, resonant
\item (not shown): labyrinth floodfill, random radius 3 kernel (7x7 disk). 
\item (ot shown): floodfill, The Circular Ruins. 
\item From Fig \ref{fig:Lighthouse}: lighthouse, concentric
\end{enumerate}

\subsection{Implementation notes.}

Eq (1) only needs FFT on a paralellizable platform. For example, in
Python using PyTorch, where $z$is the state, $I$ the input, and
$Ut$ the Fourier transform of the kernel $U$, all three torch.tensor()s
of the same shape. 
\begin{verbatim}
import torch
def iterate():     
    global Ut, z, I 
    z=torch.fft.ifft(U * torch.fft.fft(z)) 
    z=z+I 
    z=z/torch.sqrt(1 + torch.abs(z)**2) 
    return 
\end{verbatim}
In Matlab it runs 
\begin{verbatim}
    z=ifft( unikernel.*fft(z) );
    z=z+I; 
    z=z./sqrt(1+abs(z).^2); 
\end{verbatim}
where the arrays can be standard or gpuArray depending on the hardware. 

\subsection{Discrete time vs. continuous time}

When shifting the description from continuous time in the form of
ordinary differential equations to discrete time in the form of a
recurrence or iteration, the natural relationship is one of exponentiation.
We review the case of traveling waves, which have recently been implicated
in storing short-term memory in RNNs \citep{waves1,waves2} and are
similarly central to our construction here. Starting with the unidirectional
half-wave equation 
\[
\frac{\partial\psi}{\partial t}=c\frac{\partial\psi}{\partial x}
\]
and remembering that the solution of a first order linear equation
$\dot{{\bf x}}=M{\bf x}$ consists of exponentiating, ${\bf x}(t)=e^{tM}{\bf x}(0)$,
we notice that the derivative operator in the right hand side is linear
and hence, the solution should be 
\[
\psi(t,x)=e^{ct\frac{\partial}{\partial x}}\psi(0,x)
\]
What is the exponential of a derivative? Expanding the exponential
in series we obtain 

\[
e^{t\frac{\partial}{\partial x}}=1+ct\frac{\partial}{\partial x}+\frac{c^{2}t^{2}}{2}\frac{\partial^{2}}{\partial x^{2}}+\frac{c^{3}t^{3}}{3!}\frac{\partial^{3}}{\partial x^{3}}+\frac{c^{4}t^{4}}{4!}\frac{\partial^{4}}{\partial x^{4}}+\cdots
\]
which when applied to $\psi(0,x)$ is nothing more than the Taylor
expansion of $\psi(0,x+ct)$, the well-known solution of the equation
above. Therefore we say that the exponential of the derivative is
a finite translation or, conversely, that the derivative is the infinitesimal
generator of translations. In our case, a derivative presents as a
real, antisymmetric component of the kernel, and hence its exponential
is an orthogonal kernel characterized by eigenvalues with absolute
value equal to 1. 

It is instructive to see how this works out in several different bases,
because of course this relationship is invariant under change of basis,
but \emph{how} it works out in different bases is surprisingly different.
The derivative is diagonal in the Fourier basis, as $d/dx\ e^{ikx}=ike^{ikx}$.
As a result the representation of the derivative is a diagonal matrix
containing the wavenumbers times $i$. More explicitly, for a discrete
Fourier transform in $N$ elements, the basis is $F_{n}=e^{2\pi i\frac{n}{N}}$
for $n=\left\{ -N/2+1:N/2\right\} $. The derivative is a diagonal
matrix $D=2\pi i{\rm diag}(\left\{ -N/2+1:N/2\right\} )$, and its
exponential $T={\rm diag}(F_{1}).$ Returning to real space, the inverse
Fourier transform of $B_{1}$ is $\delta_{i1}$, the vector which
is $1$ on element $1$ and zero otherwise, and the convolution of
$\delta_{i1}$ with an arbitrary vector is the $+1$ circular shift. 

\medskip{}
\begin{table}[t]
\begin{verbatim}
In [1]: from numpy import *
In [2]: from scipy.linalg import *
In [3]: expm(diag(arange(1,9),1)) 
Out[3]:  
array([[ 1.,  1.,  1.,  1.,  1.,  1.,  1.,  1.,  1.],
       [ 0.,  1.,  2.,  3.,  4.,  5.,  6.,  7.,  8.],
       [ 0.,  0.,  1.,  3.,  6., 10., 15., 21., 28.],
       [ 0.,  0.,  0.,  1.,  4., 10., 20., 35., 56.],
       [ 0.,  0.,  0.,  0.,  1.,  5., 15., 35., 70.],
       [ 0.,  0.,  0.,  0.,  0.,  1.,  6., 21., 56.],
       [ 0.,  0.,  0.,  0.,  0.,  0.,  1.,  7., 28.],
       [ 0.,  0.,  0.,  0.,  0.,  0.,  0.,  1.,  8.],
       [ 0.,  0.,  0.,  0.,  0.,  0.,  0.,  0.,  1.]])

\end{verbatim}
\caption{Generating Pascal's triangle as a matrix exponential}

\end{table}

\begin{verbatim}

\end{verbatim}
Consider now the space of polynomials of degree $D$, which is a $\left(D+1\right)$-dimensional
vector space. A natural basis is that of monomials, i.e. $1,x,x^{2},x^{3},\cdots,x^{D}$.
In this basis, the derivative has as a representation the matrix given
by $D=diag(1:D,1)$ : the first above-center diagonal contains the
numbers $1,2,\cdots,D$ --- in other words, the derivative of a monomial
$x^{k}$ is $k$ times the \emph{previous} monomial $x^{k-1}$. The
exponential $T=e^{D}$ is, in fact, Pascal's triangle, the structure
that tells you the projection of $(x+1)^{k}$ onto the basis. 

Therefore, $e^{D}x^{3}=1+3x+3x^{2}+x^{3}=(x+1)^{3}$ and in general
$e^{D}P(x)=P(x+1)$
\end{document}